# Effect of Halide Composition on the Photochemical Stability of Perovskite Photovoltaic Materials

*Ravi K. Misra[§,a,b] Laura Ciammaruchi[§,a] Sigalit Aharon,[b] Dmitry Mogilyansky,[c] Lioz Etgar,[b] Iris Visoly-Fisher*[a,c] and Eugene A. Katz[*a,c]*

a. Dept. of Solar Energy and Environmental Physics, Swiss Institute for Dryland Environmental and Energy Research, The Jacob Blaustein Institutes for Desert Research (BIDR), Ben-Gurion University of the Negev, Sede Boqer Campus 84990, Israel

b. Casali Center for Applied Chemistry, The Institute of Chemistry, The Hebrew University of Jerusalem, Jerusalem 91904, Israel,

c. Ilse Katz Institute for Nanoscale Science & Technology, Ben-Gurion University of the Negev, Be'er Sheva 84105, Israel

[§]These authors contributed equally to this work.
*irisvf@bgu.ac.il, keugene@bgu.ac.il

**ABSTRACT:** Photochemical stability of encapsulated films of mixed halide perovskites with a range of $MAPb(I_{1-x}Br_x)_3$ compositions (solid solutions) was investigated under accelerated stressing using concentrated sunlight. Evolution of light absorption and the corresponding structural modifications in the films were recorded by UV-vis spectroscopy and X-ray diffraction (XRD), respectively. $MAPbBr_3$ film exhibited no degradation. In $MAPbI_3$ and mixed halide $MAPb(I_{1-x}Br_x)_3$ films, decomposition of the perovskite material with crystallization of inorganic $PbI_2$, and the corresponding degradation of light absorption, were recorded. Introduction of bromine into the solid solution was found to stress its structure and accelerate the degradation. The relevance of accelerated testing to standard operational conditions of solar cells was confirmed by comparison to degradation experiments under outdoor sunlight exposure. Reasons for the reduced stability of the mixed halide compositions are discussed.

**TOC GRAPHICS:**

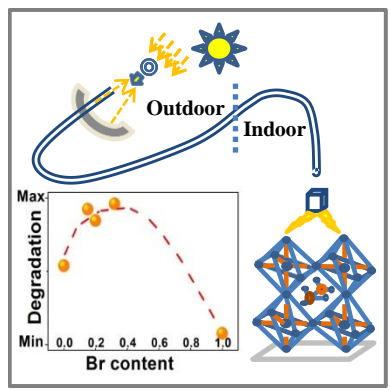

Organic−inorganic hybrid perovskites have recently achieved a breakthrough in the field of thin film photovoltaics (PV), with extremely high power conversion efficiencies (PCE) of over 20% for wet-processed devices.[1] The structure of such perovskites is $AMX_3$, where M is a divalent cation, X are halides, and A is an organic cation. Such perovskites have excellent properties for PV sunlight conversion, including direct band gaps, large absorption coefficients and high carrier mobility.[2] The ability to tune the energy band gap by changing the halide anion composition is another important advantage of these materials.[3] Seok et al. showed colorful, efficient perovskite solar cells using different I/Br ratios with PCE of up to 12.3%.[4] Etgar et al. used a two-step deposition technique to synthesize $CH_3NH_3Pb(I_{1-x}Br_x)_3$ ($MAPb(I_{1-x}Br_x)_3$, $0 \leq x \leq 1$) perovskites and utilized them in hole conductor free PV cells that combined the perovskites' good hole conduction and tunable absorption spectrum.[5]

A significant challenge en route to commercialization of perovskite-based PV is the development of devices combining high PCE and operational stability. Accelerated stability tests[6] can rapidly screen different PV materials and enhance the development of stable and efficient devices. Recently we demonstrated such accelerated stability study of perovskite thin films using concentrated sunlight.[7] In particular, encapsulated $MAPbBr_3$ films were shown to be more stable than $MAPbI_3$ films following exposure to concentrated sunlight of 100 suns (1 sun = 100 mW/cm$^2$), with the latter decomposing into $PbI_2$ and organic components. We then postulated that combining the better photovoltaic performance of $MAPbI_3$ with the added stability of its bromide counterpart, by using mixed compositions of $MAPb(I_{1-x}Br_x)_3$, would yield an optimized PV material.[7] Seok et al. demonstrated enhanced "shelf-life" stability of the PCE of cells based on $MAPb(I_{1-x}Br_x)_3$ with *x* ranging between 0.2 and 0.29, after cell storage in the dark in controlled humidity atmosphere.[4] Han et al., in their review article, attributed this improved stability to the more compact and dense crystal structure of the mixed halide perovskites.[8] On a different note, reversible instability phenomena were demonstrated in relatively Br-rich compositions: McGehee et al.[9] noted a change in photoluminescence in films with x ≥ 0.2 under visible light soaking, which was attributed to segregation into iodide-rich and bromide-rich domains. This change was fully reversible after few minutes in the dark, hence it cannot be related to long-term degradation. Friend et al. noticed the possible existence of two phases in fresh mixed halide films with x = 0.4 and 0.6, which evolved into a single phase after a few days of storage at room temperature, with no further changes.[10] Though both phenomena were short-termed, they may point to a certain instability of mixed halide films at specific compositions.

Herein we compare the photochemical stability of encapsulated films of mixed halide perovskites with a range of MAPb($I_{1-x}Br_x$)$_3$ compositions under accelerated stressing tests using concentrated natural sunlight at 100 suns. The results are compared to degradation under outdoor sunlight exposure to establish the relevance of the accelerated testing to standard operational conditions of PV cells. The evolution of light absorption and the corresponding structural modifications in the films following stressing were investigated. Contrary to our initial presumption, we found that MAPbI$_3$ and MAPbBr$_3$ are more stable than mixed halide compositions, under both concentrated and standard solar exposures. The reasons for the reduced stability of the mixed halide compositions are discussed.

The film structure and composition were studied using UV-vis absorbance spectroscopy and X-ray diffraction (XRD). Figures 1 and 2 show the XRD patterns of MAPb($I_{1-x}Br_x$)$_3$ films of various compositions, and their properties are summarized in Table 1. XRD patterns of MAPbI$_3$ and MAPbBr$_3$ films for the entire 2$\theta$ ranges (Figures 1a, e) indicate their tetragonal *I*4/*mcm* and $Pm\bar{3}m$ cubic perovskite structures, respectively.[4, 11] For pure MAPbI$_3$ (bottom curve in Figure 2), two diffraction peaks located at 28.11° and 28.36° (bottom curve in Figure 2) are associated with the (004) and (220) planes in the tetragonal *I4/mcm* phase.[12] In accordance with a recent report,[4] increasing *x* was found to result in: (a) a pattern shift to larger scattering angles (Figure 2) and the corresponding decrease of the lattice parameters (Table 1); (b) merging of the (004) and (220) peaks to a single peak at 28.5°. Finally, the XRD pattern of the pure MAPbBr$_3$ (upper curve in Figure 2) exhibited a narrow peak at 30.1°, associated with the (200) plane in the $Pm\bar{3}m$ cubic phase.

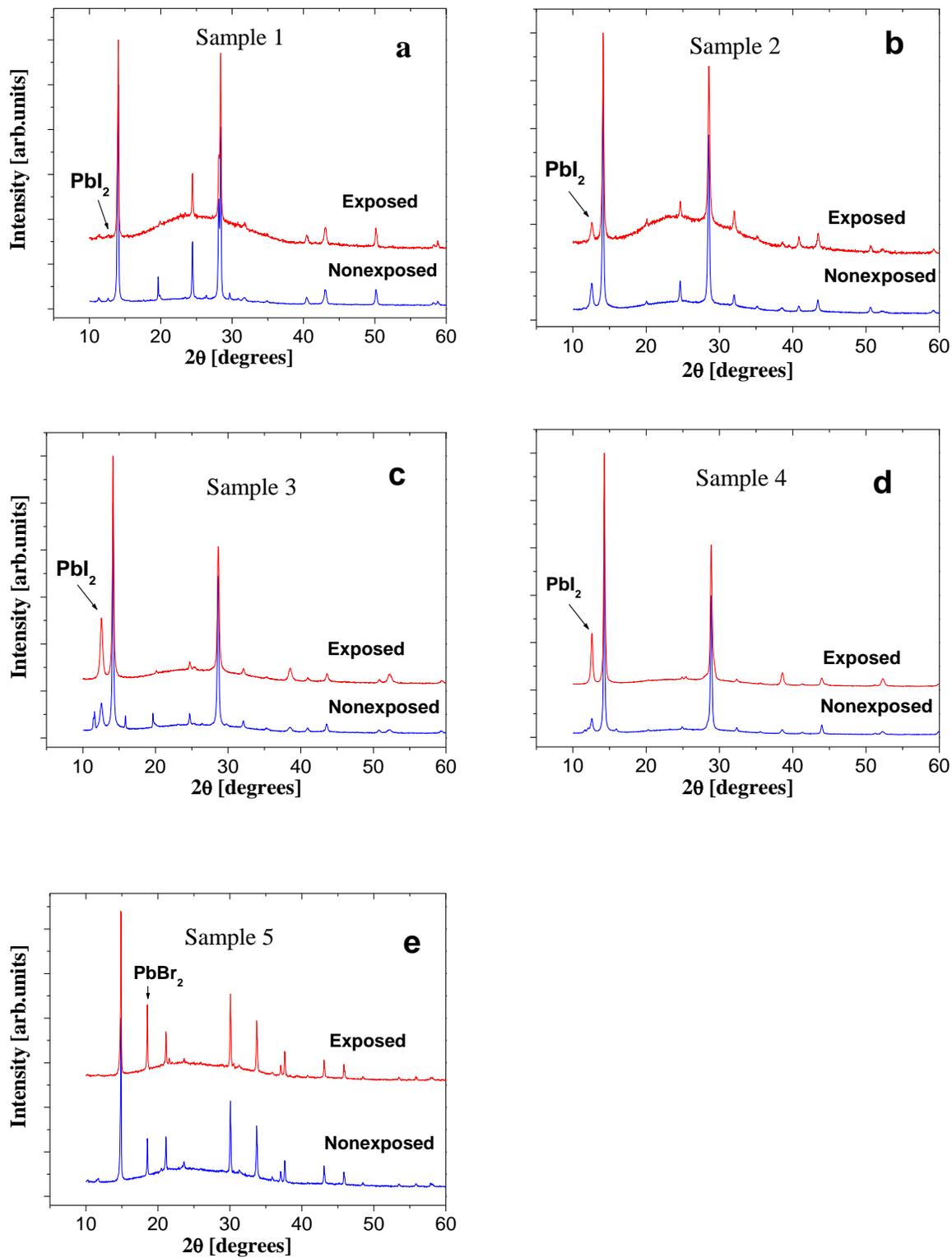

**Figure 1.** XRD patterns of MAPb($I_{1-x}Br_x$)$_3$ films before and after exposure to concentrated sunlight of 100 suns for 1 hour. Increasing sample numbers indicate increasing Br content; for sample details see Table 1.

**Table 1.** The halide compositions and structural parameters of MAPb(I$_{1-x}$Br$_x$)$_3$ thin-films.

| Sample number | MAI:MABr ratio in the organic precursor solution | Lattice parameter (Å) | Full Width at Half Max., FWHM** [deg] | Expected Br content ($x$) in the film (according to the precursor materials) | Br content ($x$) calculated from the XRD data assuming Vegard's law | *Optical bandgap* ($E_g$, eV) determined from UV-vis spect. | Br content ($x$) estimated using empirical quadratic relation for $E_g$ from Ref. 4 |
|---|---|---|---|---|---|---|---|
| 1 | 1:0 | 6.312* | 0.16 | 0 | 0 | 1.55 | 0 |
| 2 | 2:1 | 6.256 | 0.228 | 0.11 | 0.15 | 1.65 | 0.178 |
| 3 | 1:1 | 6.236 | 0.223 | 0.16 | 0.20 | 1.69 | 0.246 |
| 4 | 1:2 | 6.186 | 0.226 | 0.22 | 0.32 | 1.78 | 0.4 |
| 5 | 0:1 | 5.934 | 0.1 | 1 | 1 | 2.3 | 1.0095 |

\* The parameter was calculated for pseudo-cubic (*pc*) lattice.[4]

\*\* The FWHM was determined for the {200} peak of the cubic lattice or averaged for {004} and {220} peaks of the tetragonal lattice (see Fig. 2).

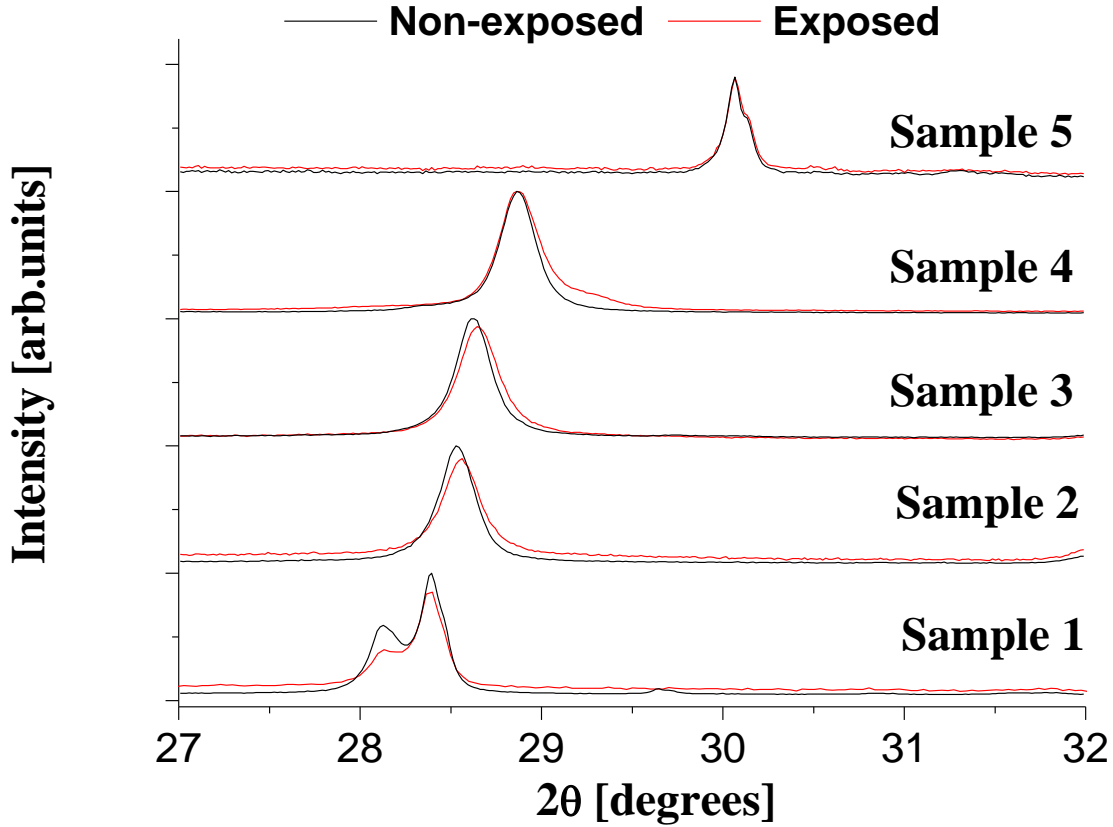

**Figure 2**. XRD patterns of MAPb(I$_{1-x}$Br$_x$)$_3$ thin-films in the 2$\theta$ range 27°-32° before and after exposure to concentrated sunlight of 100 suns for 1 hour. Increasing sample numbers indicate increasing Br content; for sample details see Table 1.

The actual halide composition of the films was determined and compared to that in the precursor solutions. According to Vegard's law, the variation of the lattice parameter in a homogeneous alloy (solid solution) is linear with composition.[13] Seok et al.[4] suggested the relevance of this rule for their MAPb(I$_{1−x}$Br$_x$)$_3$ films at the entire range of x (0 ≤ x ≤ 1). Applying Vegard's law to our XRD results showed a considerable difference between the bromine content between the expected and actual composition (Table 1). The Br content was also determined from the shift of the fundamental absorption edge recorded by UV-vis absorbance spectroscopy.[4,14] Figures 3 and S1 show UV-vis light absorbance spectra of the various MAPb($I_{1-x}Br_x$)$_3$ films. Tauc plots were used to determine the band gap ($E_g$) values (insets in Figure 3).[14, 15] A gradual increase of the $E_g$ values with increasing Br content is in evidence (Table 1). The $E_g$ values for MAPbI$_3$ and MAPbBr$_3$ samples are in agreement with previously reported ones, ~1.58[16] and 2.3 eV,[17] respectively. Application of the empirical quadratic relation $E_g(x) = 1.57 + 0.39x + 0.33x^2$, postulated in Ref. 4 for MAPb($I_{1-x}Br_x$)$_3$ mixture films and used in a number of subsequent studies,[14, 18] suggests values of Br concentration in the films in agreement with those obtained from XRD analysis using Vegard's law (Table 1). We conclude that the mixed halide films have a single phase structure (solid solution) with Br content slightly larger than that expected from the precursor materials, probably due to the sequential, two-step deposition method (see Experimental Methods). The relationship between the halide composition of the precursor solution and the films can be strongly dependent on the technology of the film preparation. Indeed, Friend et al. confirmed that the ratios of bromine to iodine in their MAPb($I_{1-x}Br_x$)$_3$ films are the same as those in the solutions since a stoichiometric precursor composition in a single deposition step was used (PbI$_2$/MAI(Br) = 1).[10] However, in other studies,[18] where an excessive amount of methylammonium salts was involved in the film preparation, a discrepancy between actual halide compositions in solid perovskites and initial precursor solutions was reported. In our case, this result can be related to the use of the two-step deposition process where PbI$_2$ or PbBr$_2$ films were dipped into isopropanol solutions containing MAI and MABr at various ratios.[5] Different reaction rates of the lead halides with MAI *vs* MABr control the actual halide composition of the films.

The evolution of light absorption and the corresponding structural modifications in the films following stressing were investigated using UV-vis absorbance spectroscopy and XRD, respectively. Figure 3a illustrates the evolution of UV-vis light absorbance spectra for the encapsulated MAPbI$_3$ film due to its exposure to concentrated sunlight of 100 suns. Concentrated sunlight accelerates stressing via increased light intensity which can be accompanied by sample heating (see Experimental Methods section). We have previously shown that the decomposition mechanism of perovskites such as MAPbI$_3$ is light- and thermally enhanced or initiated, while heat alone did not induce decomposition.[7] Following the first 20 minutes of exposure the absorption decreased in the range of ~520-620 nm forming a hump around 520 nm. With increased exposure time, this decrease expands to a wide range of 520-800 nm and the hump becomes more prominent at ~500 nm. The overall degradation in absorption of this film was quantified by calculating the Absorption Degradation State (ADS), i. e., the ratio of the number of absorbed solar photons (from the AM1.5G spectrum) by the exposed film to that by the fresh sample[7] (for the calculation method see Supporting Information). This ratio showed a slight gradual decrease with the exposure time (by ~2% for 60 minutes, maximum dose of 100 sun*hours, black line in Figure 4a).

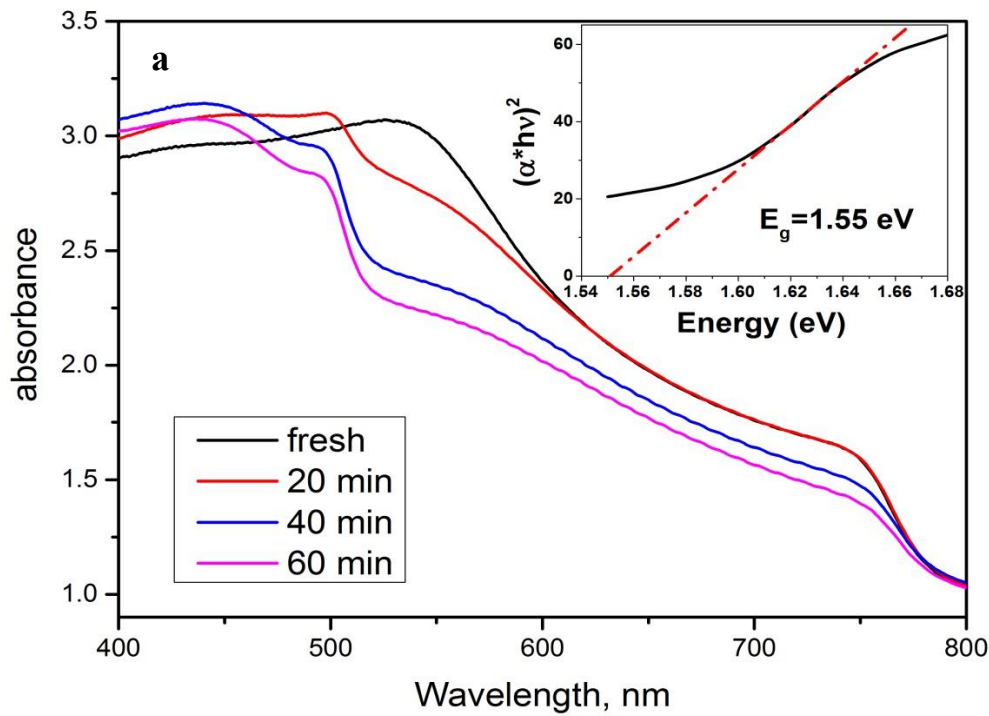

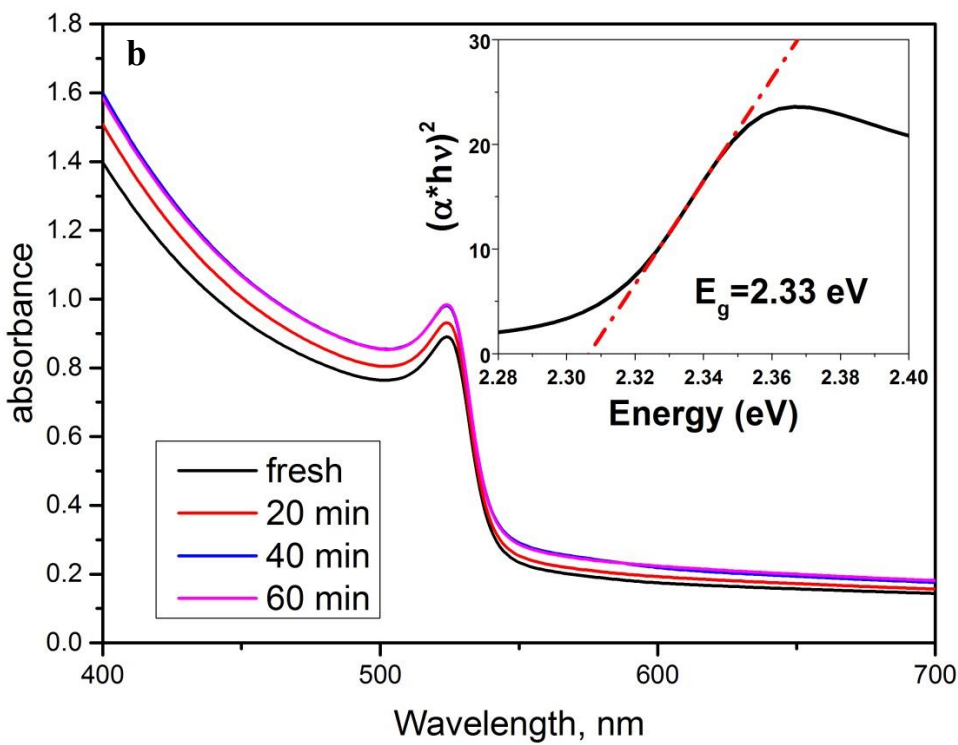

**Figure 3.** Evolution of UV-vis light absorbance spectra for (a) MAPbI$_3$ (sample number 1) and (b) MAPbBr$_3$ (sample number 5) encapsulated films following exposure to concentrated sunlight of 100 suns. The legends indicate the exposure times. The insets show the spectrum of the fresh film near the fundamental absorption edge plotted in Tauc's coordinates.[15]

The evolution of light absorption of the mixed halide $MAPb(I_{1-x}Br_x)_3$ films showed similar trends (Figure S1) but at considerably higher degradation rates (green, blue and purple curves in Figure 4a). The largest decrease in ADS (by ~ 13 %) was shown by sample number 4, the most Br-rich mixed halide film. The same message is provided by the black curve in Figure 5 showing the normalized difference in absorption following degradation (1-ADS) as a function of Br content in the film, $x$, calculated using the XRD analysis (Table 1). At the same time UV-vis absorbance spectrum of pure $MAPbBr_3$ did not exhibit any degradation following exposure (Fig. 3b), but rather a slight improvement in the absorption (see also increase in ADS for this film in Figure 4 and negative final "degradation parameter" (1-ADS) = -0.1 in Figure 5). We have verified that this increased absorption cannot be explained by degradation, as $PbBr_2$ has negligible absorption in the range of 400-800 nm (Fig. S2). The slight increase in absorption might be explained by enhanced conversion of the $PbBr_2$/ MABr precursors to the final $MAPbBr_3$ perovskite in the film. This conversion was not complete in the pristine film, as indicated by the presence of $PbBr_2$ detected by XRD (Fig. 1e). However, this conversion was too small to be reflected in the XRD after sunlight exposure. This effect should be further investigated.

Joint analysis of the light absorbance (Figures 3a and S1) and XRD data (Figure 1) suggests that decomposition of the perovskite via crystallization of $PbI_2$ occurs during exposure of $MAPbI_3$ film and mixed halide $MAPb(I_{1-x}Br_x)_3$ films, in accordance with our recent results.[7] Decomposition of $MAPbI_3$ to $PbI_2$ and MAI was previously suggested as the result of photolysis, with further dissociation of MAI to methyl amine, iodine and other small molecules (hydrogen, HI, water).[19-21] $PbI_2$ crystallization was directly confirmed herein by XRD measurements (Figure 1). The observed shift of the XRD perovskite-related peaks towards higher $2\theta$ values following exposure (Figure 2) indicated an increased Br content in the remaining perovskites following $PbI_2$ release. The lack of splitting of these perovskite-related XRD peak suggests no phase separation to the $MAPbI_3$ and $MAPbBr_3$ perovskites. FTIR measurements showed a decrease in organic (carbon- and nitrogen- related) groups' absorption, with no alternative peaks rising (Fig. S3), indicating their elimination from the solid film. Electron paramagnetic resonance spectroscopy of our perovskite samples illuminated by 100 suns in a closed ampoule showed the involvement of MAI dissociation in the degradation process (to be published elsewhere). We therefore suggest a photolysis decomposition of the mixed halide perovskite to MAI, $PbI_2$ and a Br-richer mixed halide perovskite as a possible first step, which is followed by further MAI decomposition into volatile species and iodine.[19-21]

To estimate the degree of the degradation, the ratio of {002} peak intensity of $PbI_2$ (indicated by arrows in Figure 1) to that of the {100} peak ($2\theta = 14°$) of I-containing MA-perovskites was calculated for the fresh and the exposed films. The ratio of the exposed films was then normalized by that of the fresh film, and the resulting values are presented in Figure 5 (blue curve). This value was found to increase with increasing $x$. While this value was almost 1 after exposure of samples 1 ($MAPbI_3$) and 2 ($MAPb(I_{1-x}Br_x)_3$, $x = 0.15$ according to the XRD analysis), it reached values of ~ 2.5 and 3.3 for larger Br contents in $MAPb(I_{1-x}Br_x)_3$ films (samples 3, 4, $x = 0.2$ and 0.32 according to the XRD analysis).

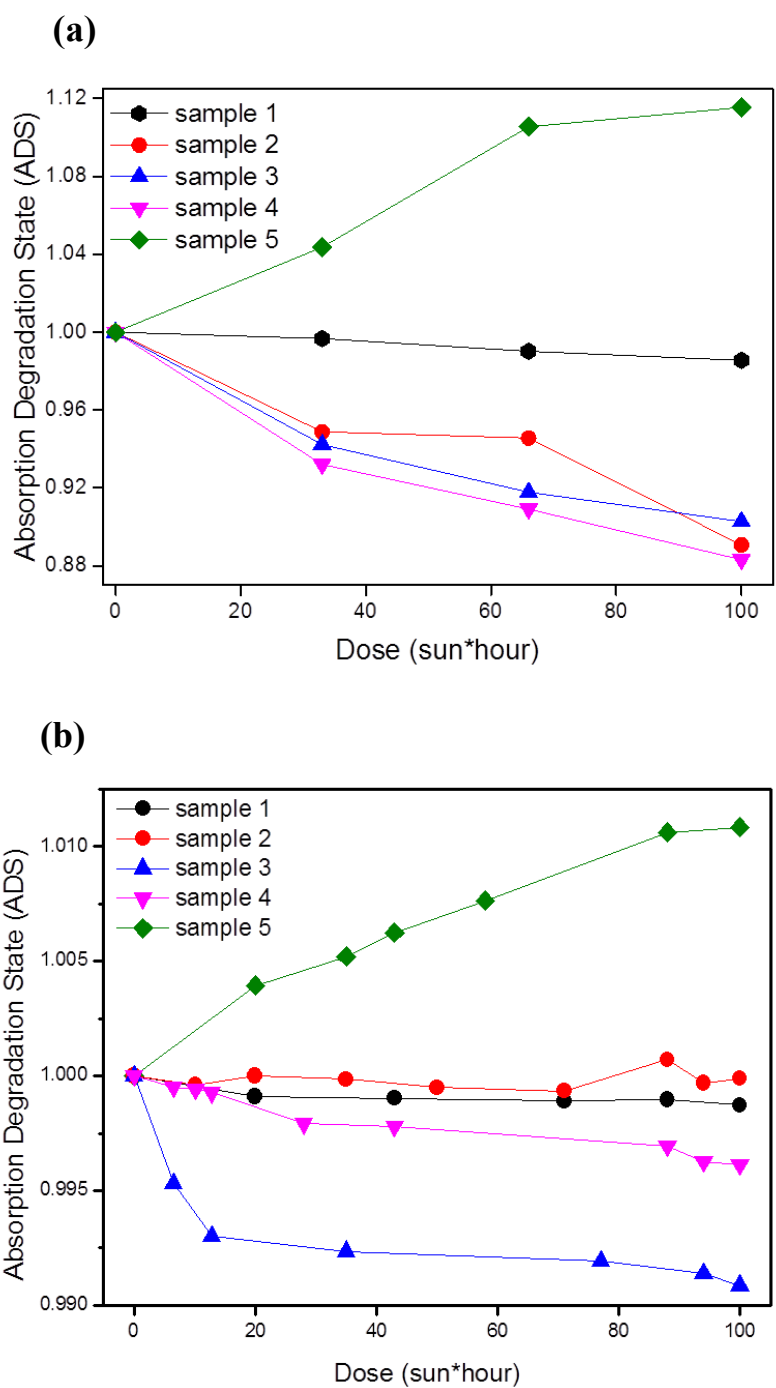

**Figure 4**. Absorption Degradation State (ADS) of MAPb($I_{1-x}Br_x$)$_3$ films as a function of the exposure dose under illumination of (a) 100 suns and (b) outdoor solar exposure. Increasing sample numbers indicate increasing Br content; for sample details see Table 1.

We conclude that, though MAPbBr$_3$ is the most stable composition, increased bromine content in the mixed halide perovskite films resulted in accelerating their photochemical degradation. We relate this finding with the observed increase in the Full Width at Half Maximum (FWHM) of the XRD peaks for mixed halide perovskites compared to MAPbI$_3$ and MAPbBr$_3$ (Fig. 5 and Table 1). Similar behavior was previously observed for CH$_3$NH$_3$Sn(I$_{1-x}$Br$_x$)$_3$ solid solutions.[22] The FWHM of XRD peaks is inversely related to the crystal coherence length, therefore the observed FWHM increase for MAPb(I$_{1-x}$Br$_x$)$_3$ indicates that the mixed halide compounds contain more structural defects and/or grain boundaries in comparison with pure halide perovskites. The decreased grain sizes in the mixed halide films (in comparison with those in MAPbI$_3$ and MAPbBr$_3$ samples) have been confirmed by Scanning Electron Microscopy (Fig. S4). Grain boundaries are thought to play a significant role in initiating photochemical degradation in perovskites.[23-24]

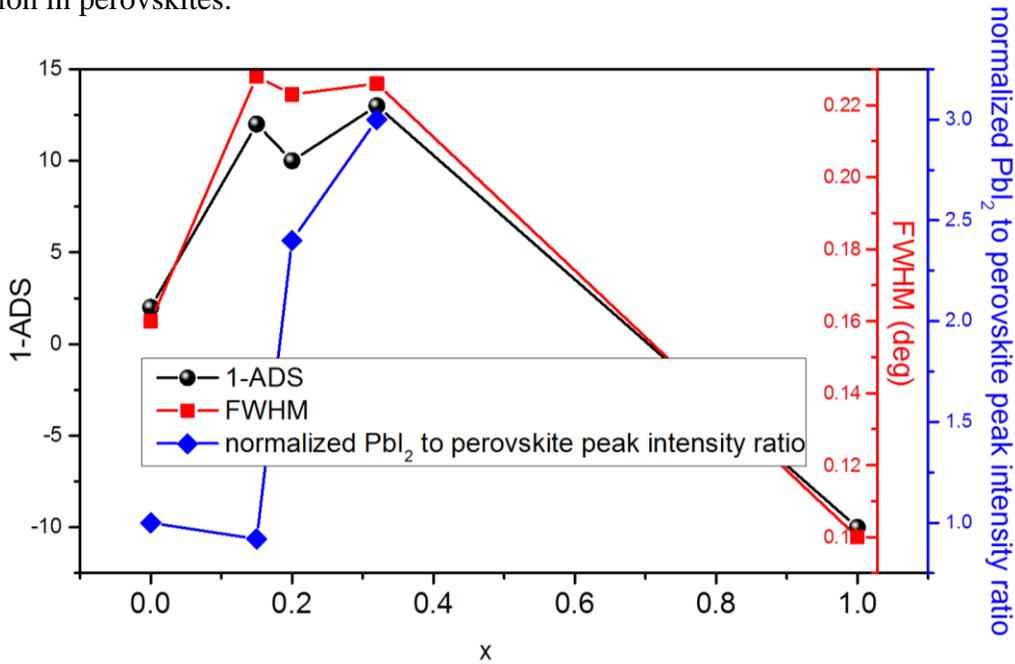

**Figure 5.** Correlation between the FWHM of the XRD peaks for fresh MAPb(I$_{1-x}$Br$_x$)$_3$ films and the degradation parameters, namely (1-ADS) and the normalized ratio between intensities of the XRD PbI$_2$ and perovskite peak intensities for these films after their exposure to concentrated

sunlight of 100 suns for 1 hr. XRD peaks used for FWHM analysis are indicted in Table 1. The x values were determined from XRD analysis.

Both the smaller grain size and the enhanced decomposition rate of the mixed perovskite films can be explained by the presence of two different halide ions with different sizes occupying 6 coordinating sites around the central Pb(II) cation. This may result in the distortion of the skeletal octahedral structure of $[PbI_6]^{4-}$ and a corresponding strain amplification in the pseudo-cubic geometry of the $MAPb(I_{1-x}Br_x)_3$ solid solutions.

To establish the relevance of accelerated testing by concentrated sunlight to standard operational conditions of PV cells, the evolution of UV-vis light absorbance spectra of $MAPb(I_{(1-x)}Br_x)_3$ films was recorded during their exposure to natural outdoor sunlight. We note that the degradation occurred only under illumination and no degradation was recorded after night-time "shelf" storage. Figure 4b shows the ADS evolution during exposure to 100 sun*hours. This accumulated dose is equivalent to the dose that samples experienced during 1 hour under 100 suns. Though the observed ADS variations are at least an order of magnitude smaller than those under the same dose of concentrated sunlight (i. e., the acceleration ratio in the concentrated sunlight experiments was larger than 100, due to the combined effect of increased light intensity and temperature[7]), the results showed the same trend of ADS evolution. Indeed, MAPbBr3 film exhibited improved ADS following exposure, while increased bromine content in the films accelerated ADS degradation.

In summary, encapsulated films of mixed halide perovskites with a range of $MAPb(I_{1-x}Br_x)_3$ compositions were prepared, and their composition and stability were characterized. The films were found to have a single phase crystalline structure with Br content larger than that expected from the precursor materials. Film exposure to concentrated sunlight of 100 suns and to outdoor 1 sun did not result in degradation of the $MAPbBr_3$ films, while $MAPbI_3$ and mixed halide $MAPb(I_{1-x}Br_x)_3$ films degraded yielding crystallization of $PbI_2$ and the corresponding degradation of light absorption. Increased bromine content accelerated this process. We relate this finding with an observed increase in the FWHM of the mixed halide perovskite XRD peaks, i.e., decreased grain sizes. The role of grain boundaries in the photochemical degradation of perovskite PV materials should be a topic of further research. We note that the mixed halide compositions studied herein are limited to maximum Br content of ~0.4. Stability studies of Br-rich compositions are currently underway.

**EXPERIMENTAL METHODS**

*Synthesis of $MAPb(I_{1-x}Br_x)_3$ perovskites and film preparation*: $CH_3NH_3I$ (MAI) and $CH_3NH_3Br$ (MABr) were synthesized as described earlier,[5] by reacting 30 mL of methylamine (40% in methanol, TCI) and 32.3 mL of hydroiodic acid (57 wt% in water, Aldrich) or 23.32 mL of hydrobromic acid (48wt% in water, Aldrich) in a 250 mL round bottom flask at $0^0C$ for 2 hours while stirring. The precipitate was recovered by putting the solution in a rotatory evaporator and carefully removing the solvents at $50^0C$ for 1 hour. The yellowish raw product of methylammonium iodide (MAI) or methylammonium bromide (MABr) was washed thrice in diethyl ether by stirring the mixture for 30 minutes. After filtration, the solid was collected and dried at $60^0C$ in a vacuum oven for 24 hours. The films of $MAPb(I_{1-x}Br_x)_3$ were prepared using the sequential deposition method. 1M solution of $PbI_2$ (Sigma Aldrich) or $PbBr_2$ (Sigma Aldrich) in dimethylformamide (Sigma Aldrich) was spin-coated onto a glass substrate (in oxygen and water free nitrogen atmosphere) at 6000 rpm for 5 seconds, followed by annealing for 5 minutes at 90 °C on a hot-plate. Then PbI$_2$ films were dipped for 30 seconds into one of the following solutions in 10 ml isopropanol (Sigma Aldrich): 0.100g MAI (sample number 1), 0.067g MAI + 0.0235g MABr (sample number 2), 0.050g MAI + 0.0352g MABr (sample number 3), 0.033g MAI + 0.047g MABr (sample number 4). PbBr$_2$ films were dipped in a solution of 0.070g MABr in 10 ml isopropanol (sample number 5). After dipping, samples were annealed for 30 minutes at 90 °C on a hot-plate. Encapsulation of these films was done inside a glovebox using a DuPont™ Surlyn® film as a spacer and sticker for the upper glass by melting it using a welding-pen.

*Film exposure to concentrated sunlight.* Sunlight collected and concentrated outdoors was focused into a transmissive optical fiber and delivered indoors onto the sample (Fig. S5a).[25,26] A 2.5 cm long, 0.25 cm$^2$ square cross-section kaleidoscope placed between the distal fiber tip and the sample was used to achieve flux uniformity (Fig. S5b).[26] Sunlight exposure was performed at Sede Boqer (Lat. 30.8°N, Lon. 34.8°E, Alt. 475 m) during clear-sky periods around noontime, where the solar spectrum was found to be very close to the standard AM1.5G solar spectrum.[27] The spectrum of light delivered to the sample was also close to the AM1.5 standard.[25] The concentration of incident sunlight on the sample was measured using a spectrally blind pyranometer (thermopile) of 5% accuracy. During the exposure, the samples were thermally bonded to the top of a thermoelectric temperature controlled plate, which was set to 25°C. The temperature at the 'sample/thermoelectric plate' interface was measured with a T type thermocouple connected to the sample using silver paste. The sample temperature during illumination was about 45-55°C.

*Film exposure to outdoor sunlight.* Solar exposure of the samples was performed during day-time in July-August 2015 with permanent monitoring of the intensity of incident sunlight and the sample temperature, while for night-time the films were kept in the dark in a glove-box. Global intensity of incident sunlight was measured with a calibrated thermopile pyranometer (Eppley PSP). The sample and the pyranometer were mounted on a fixed angle stand (30° to horizontal, see for example Figure 6c in Ref. 28). The samples experienced ambient temperatures (20-25°C) measured by thermocouples.

*Optical and crystallographic characterization.* The UV-vis absorbance spectra of the films before and after exposure were recorded in direct transmission mode using a Cary 5000 UV-vis-NIR spectrophotometer (Agilent technologies). To verify that diffused scattering was not responsible for the observed changes in the measured absorbance, the diffuse scattering intensity spectra were measured after exposure and were found to be almost entirely independent of the sample's light exposure (Fig. S6). The absorbance is defined as $A=\varepsilon cl$, where $\varepsilon$ is the material's attenuation coefficient, $c$ is the concentration of the absorbing species in the film, and $l$ is the film thickness. Hence changes in $A$ reflect the changes in the concentration of the absorbing species in the film, $c$. X-ray Diffraction (XRD) data were collected with Panalytical Empyrean Powder Diffractometer equipped with position sensitive X'Celerator detector using Cu $K_\alpha$ radiation ($\lambda$=1.5405 Å) and operated at 40 kV and 30 mA.

**ASSOCIATED CONTENT**

**Supporting Information.** Schematic representation of the solar concentrator, UV-Vis absorption spectra for the mixed halide $MAPb(I_{1-x}Br_x)_3$ perovskite films, $PbBr_2$ absorbance spectrum, FTIR spectra of the $MAPbI_3$ film before and after exposure, SEM images, diffuse reflectance spectra, and the calculation method of the number of absorbed solar photons are available as additional supporting information. This material is available free of charge via the Internet at http://pubs.acs.org.

**AUTHOR INFORMATION**

**Corresponding Author**

*Email: keugene@bgu.ac.il

**Notes**

The authors declare no competing financial interests.

**ACKNOWLEDGMENTS**

R.K.M., E.A.K. and I.V.F. are members of the European Commission's StableNextSol COST Action MP1307. R.K.M. is thankful to BGU's Blaustein Center for Scientific Cooperation for a postdoctoral research fellowship. The research was funded in part by the Adelis Foundation. L. E. thanks the German Israel Foundation for its financial support for Young Researchers. L.C. is thankful to ENEA (Ente Nazionale Energia e Ambiente) and the Italian Ministry of Foreign Affairs for a visitor fellowship. E. A. K. acknowledges Dr. P. A. Troshin for a fruitful discussion.

# Supporting Information

# Effect of Halide Composition on the Photochemical Stability of Perovskite Photovoltaic Materials

*Ravi K. Misra[§,a,b] Laura Ciammaruchi[§,a] Sigalit Aharon,[b] Dmitry Mogilyansky,[c] Lioz Etgar,[b] Iris Visoly-Fisher\*[a,c] and Eugene A. Katz[\*a,c]*

a. Dept. of Solar Energy and Environmental Physics, Swiss Institute for Dryland Environmental and Energy Research, The Jacob Blaustein Institutes for Desert Research (BIDR), Ben-Gurion University of the Negev, Sede Boqer Campus 84990, Israel

b. Casali Center for Applied Chemistry, The Institute of Chemistry, The Hebrew University of Jerusalem, Jerusalem 91904, Israel,

c. Ilse Katz Institute for Nanoscale Science & Technology, Ben-Gurion University of the Negev, Be'er Sheva 84105, Israel

[§]These authors contributed equally to this work.

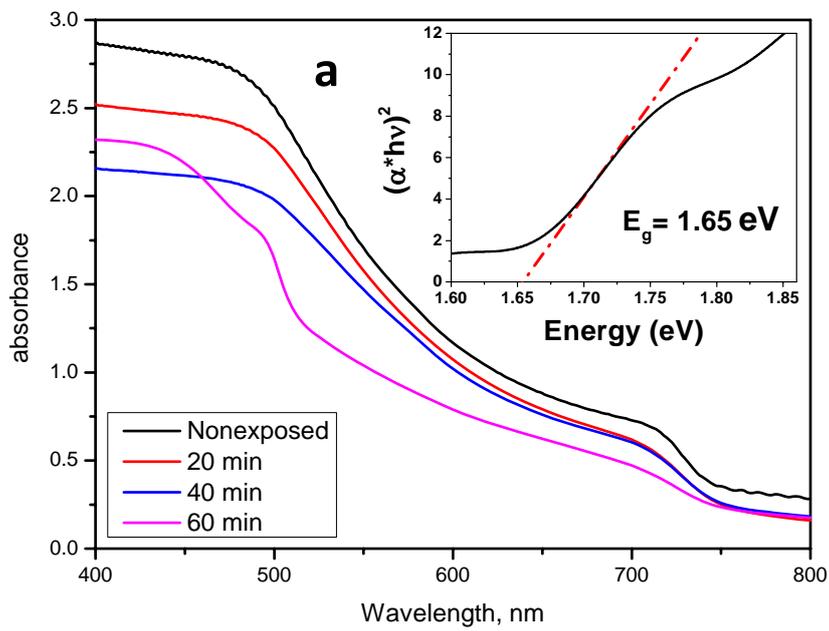

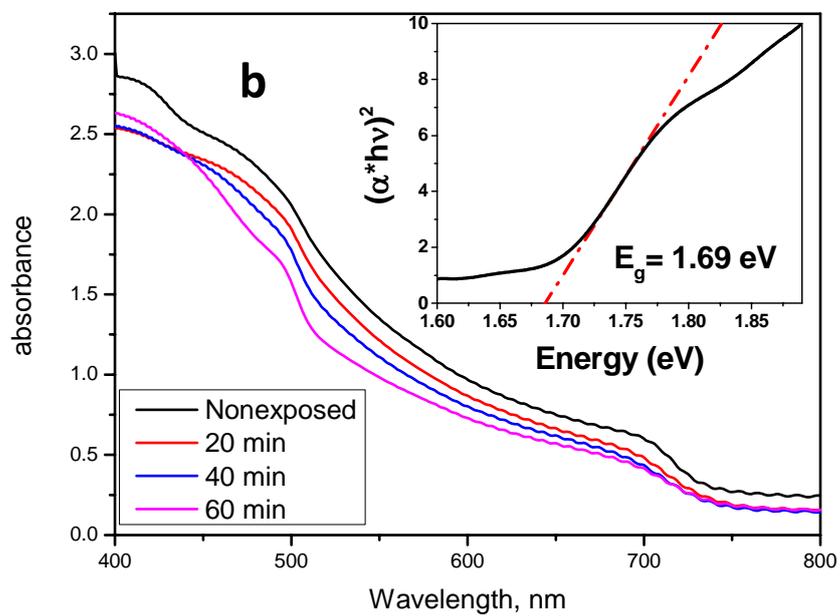

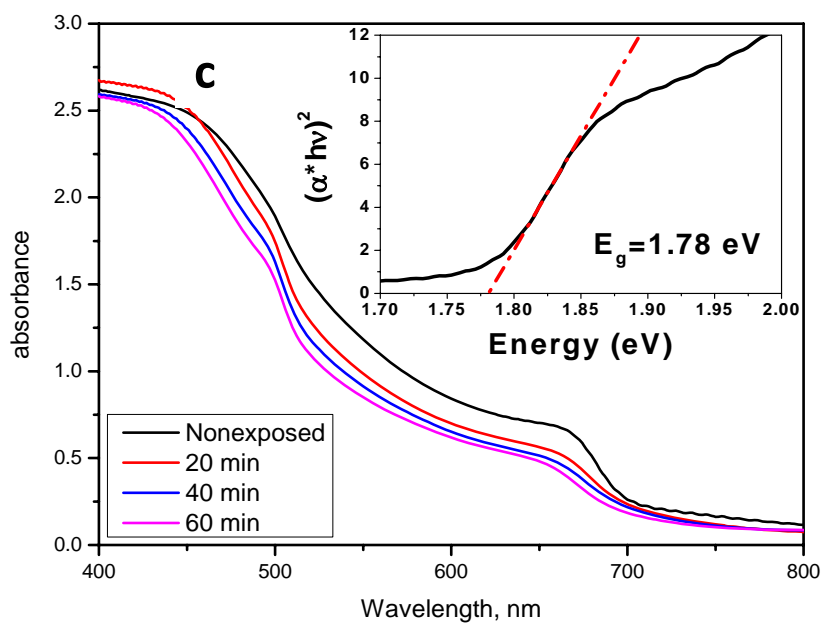

**Figure S1.** Evolution of UV-vis light absorption spectra for MAPb(I$_{1-x}$Br$_x$)$_3$ films following exposure to concentrated sunlight of 100 suns: (a) sample 2, (b) sample 3, (c) sample 4. Increasing sample numbers indicate increasing Br content; for sample details see Table 1. The legends indicate exposure times. The insets show Tacu plots of the fresh film absorption near the fundamental absorption edge.

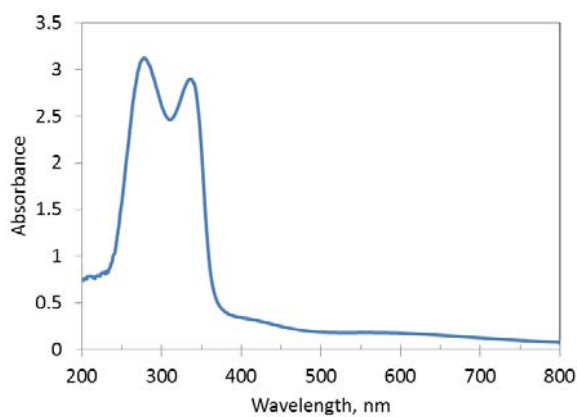

**Figure S2.** PbBr$_2$ absorbance spectrum.

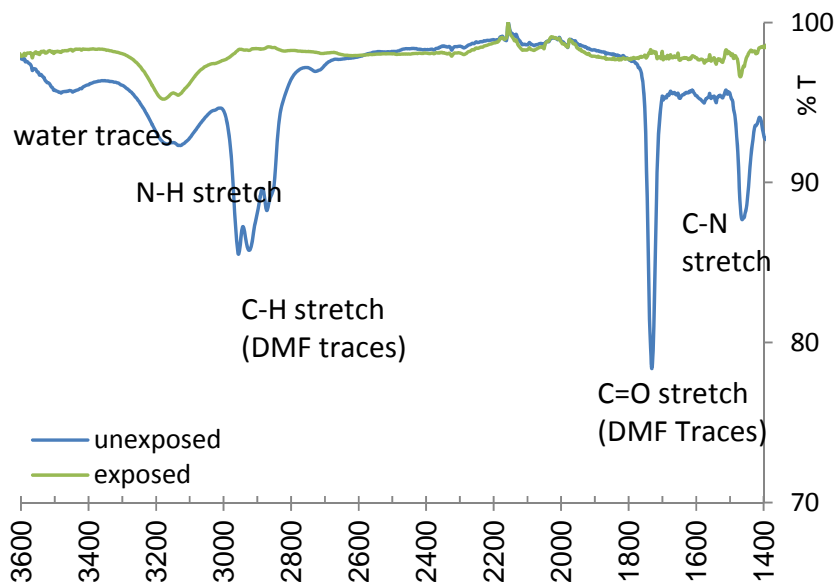

**Figure S3.** FTIR spectra of the MAPbI$_3$ film (sample 1) before and after exposure to 100 suns*hours. Measurements were performed using a Thermo Scientific Nicolet 8700 FTIR spectrometer with a deuterated triglycine sulfate (DTGS) detector. The samples were prepared by grinding the film and mixing it with KBr powder at a concentration of 0.5% w/w, followed by pressing into a pellet. Peak identification was based on previously published data [1].

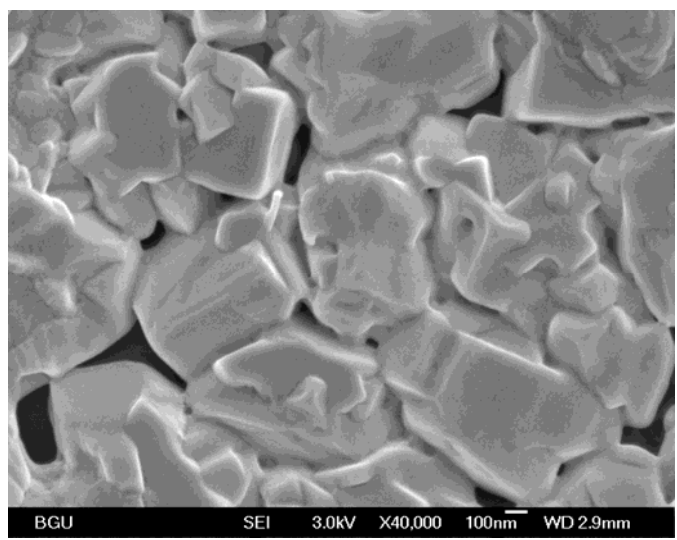
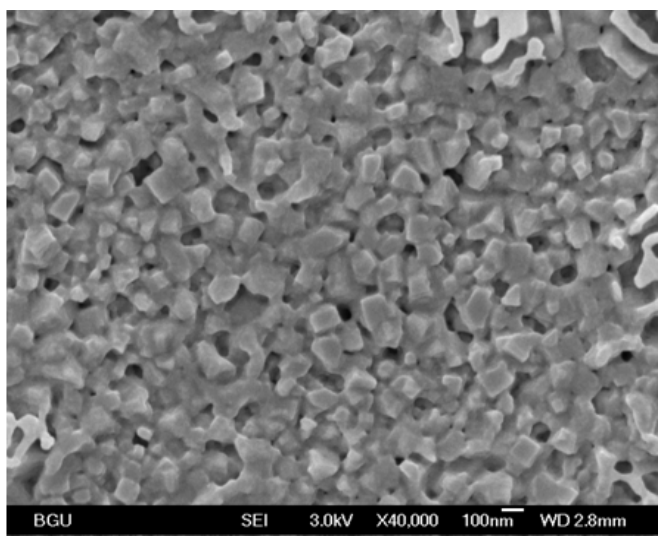

**a**  **b**

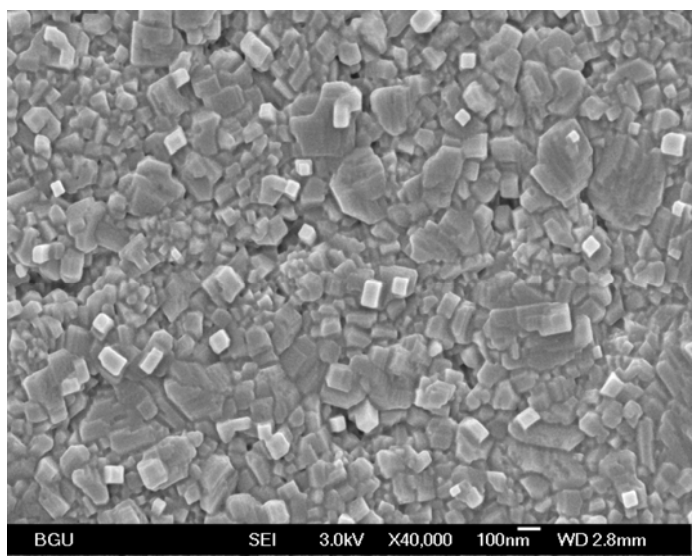

**c**

**Figure S4.** SEM images of perovskite films: (a) sample 1; (b) sample 3; (c) sample 5. The images show that MAPbI$_3$ (sample 1) has the largest grains, MAPb(I$_{1-x}$Br$_x$)$_3$ (x=0.2 according to XRD analysis, sample 3) has the smallest grains, and MAPbBr$_3$ (sample 5) has a mixture of small and large grains. Images were taken using a JEOL 7400F field emission gun high resolution scanning electron microscope (HRSEM) in SE mode using an in-lens detector.

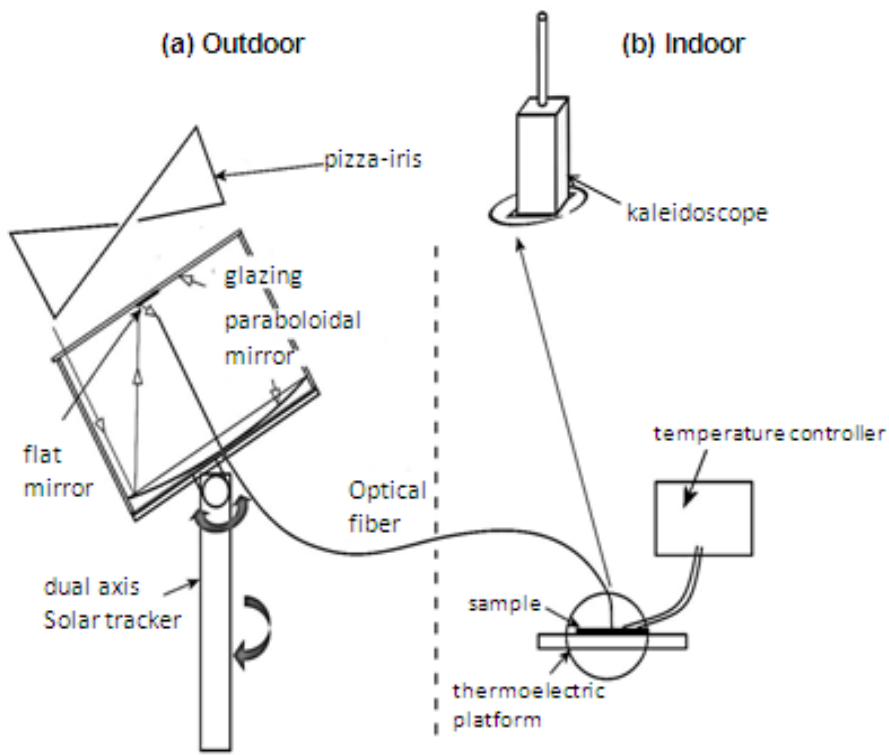

**Figure S5.** Schematic description of the Minidish dual-axis tracking solar concentrator (20 cm in diameter): (a) The outdoor set-up concentrates the solar radiation at the tip of a highly transmitting optical fiber, which guides the concentrated sunlight indoors onto the sample being tested, with (b) a uniform irradiation of the sample via a kaleidoscope [2-3].

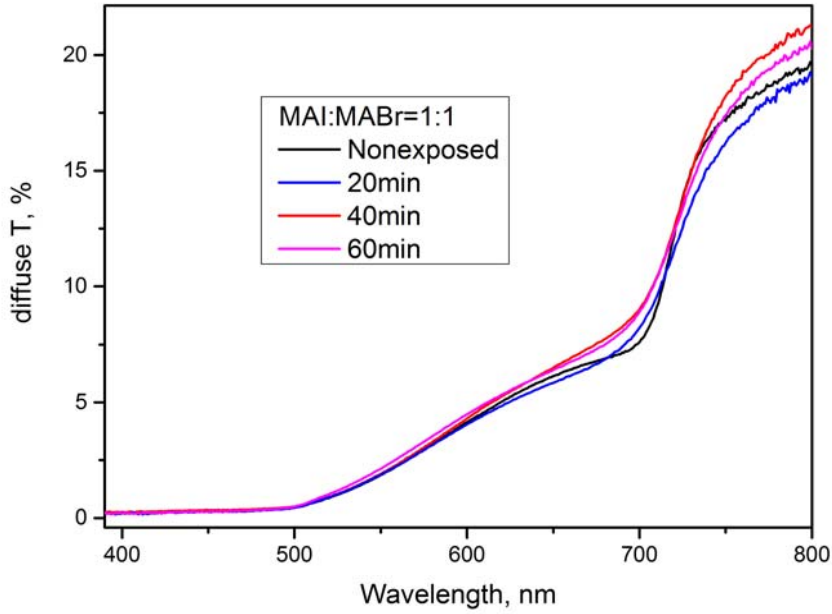

**Figure S6.** The diffuse scattering intensity of MAPb(I$_{1-x}$Br$_x$)$_3$ film (x=0.2 according to XRD analysis, sample 3) following different exposure times to 100 suns.

Calculating the number of absorbed solar photons:

To quantify the degradation in absorption, the number of absorbed solar photons ($N_{Tot}^t$) was calculated as a function of exposure time. For each wavelength in the range 400-800 nm, the percentage of absorbed photons was calculated from the absorbance spectra, and multiplied by the number of incident solar photons. The resulting number of absorbed photons was summed over the wavelength range 400-800 nm providing the total number of absorbed solar photons $N_{Tot}^t$:

$$N_{Tot}^t = \sum_{\lambda 1}^{\lambda 2} N_0(\lambda)(1 - 10^{-A^t(\lambda)})$$

A$^t$(λ) - the measured absorbance at a given wavelength λ and after exposure time t, $N_0(\lambda)$ - the incident photon flux, λ1=400 nm and λ2=800 nm. A$^t$(λ) was directly extracted from the UV–vis absorbance spectra after the corresponding exposure time, and the photonic flux was taken from the ASTMG173 standard, which was used as the AM1.5G reference spectrum [4]. To follow the evolution of the degradation of the perovskite, $N_{Tot}^t$ was normalized by the number of photons absorbed by the fresh sample $N^o_{tot}$, to provide the ratio indicated as "absorption degradation state" (ADS).